\documentclass[12pt]{article}
\def\beq{\begin{equation}}
\def\eeq{\end{equation}}
\def\bea{\begin{eqnarray}}
\def\eea{\end{eqnarray}}
\def\bef{\begin{figure}}
\def\enf{\end{figure}}
\def\S{{\bf S}}
\def\C{{\bf C}}
\def\Z{{\bf Z}}
\def\R{{\bf R}}
\def\P{{\bf P}}

\def\G{{\bf G}}
\def\CC{{\cal C}}
\def\CD{{\cal D}}

\def\CN{{\cal N}}
\def\CO{{\cal O}}

\def\We{{W_{\mbox{eff}}}}

\def\ba{\begin{array}}
\def\ea{\end{array}}
\def\bce{\begin{center}}
\def\ece{\end{center}}

\def\del{\partial}

\def\IC{{\relax\hbox{$\inbar\kern-.3em{\rm C}$}}}
\def\ID{\relax{\rm I\kern-.18em D}}
\def\IE{\relax{\rm I\kern-.18em E}}
\def\IF{\relax{\rm I\kern-.18em F}}
\def\IG{\relax\hbox{$\inbar\kern-.3em{\rm G}$}}
\def\IGa{\relax\hbox{${\rm I}\kern-.18em\Gamma$}}
\def\IH{\relax{\rm I\kern-.18em H}}
\def\II{\relax{\rm I\kern-.18em I}}
\def\IK{\relax{\rm I\kern-.18em K}}

\def\IQ{\relax\hbox{$\inbar\kern-.3em{\rm Q}$}}

\begin{document}
\begin{titlepage}
\rightline{HUTP-01/A022} \rightline{HU-EP-01/20} \rightline{hep-th/0105066}
\def\today{\ifcase\month\or
January\or February\or March\or April\or May\or June\or July\or August\or 
September\or October\or November\or
December\fi, \number\year} \vskip 1cm \centerline{\Large \bf Geometric
Transition, Large N Dualities}\centerline {\Large \bf and MQCD Dynamics} \vskip 
1mm
\vskip 1cm \centerline{\sc Keshav Dasgupta $^{a,}$\footnote{keshav@ias.edu}, 
Kyungho
Oh$^{b,}$\footnote{On leave from Dept. of Mathematics, University of 
Missouri-St. Louis, oh@hamilton.harvard.edu}
and Radu Tatar$^{c,}$\footnote{tatar@physik.hu-berlin.de}} \vskip 1cm 
\centerline{{ \it $^a$ School of Natural
Sciences, Institute for Advanced Study, Princeton NJ 08540, USA}} \vskip 1mm 
\centerline{{ \it $^b$ Lyman
Laboratory of Physics, Harvard University, Cambridge, MA 02138, USA}} \vskip 1mm 
\centerline{{\it $^c$ Institut
fur Physik, Humboldt University, Berlin, 10115, Germany}} \vskip 2cm 
\centerline{\sc Abstract} \vskip 0.2in
We study Vafa's geometric transition from a brane setup in M-theory. In this 
transition D5
branes wrapped on $\P^1$ cycles of a resolved conifold
disappear and are replaced by fluxes on a deformed conifold. In the limit
of small sized $\P^1$, we describe this mechanism as 
a transition from curved M5 branes to plane
M5 branes which replaces $SU(N)$ MQCD by $U(1)$ theories on the bulk.
 This  agrees with the results expected from the geometric transition.
We also discuss the reduction to ten dimensions and a brane 
creation mechanism in the presence of fluxes.

\vskip 1in \leftline{May 2001}
\end{titlepage}
\newpage

\section{Introduction}
Many interesting results have been obtained by realizing gauge field theories on 
the world-volume of branes in
string theory (for a review see \cite{gk}). One of the most interesting 
direction was to study aspects of
confinement in the M theory fivebrane version of QCD (MQCD) 
\cite{wit,WittenMQCD}. The brane theory (MQCD) is not
identical to QCD as it contains the Kaluza-Klein modes from the compactified
direction $x^{10}$ and the
radius of the direction $x^{10}$ becomes an
 extra parameter. In \cite{WittenMQCD}, different aspects
of ${\cal N} = 1$ MQCD were studied, it was shown to have flux tubes and the 
tensions of MQCD strings and domain walls
were derived.

One question which arises is the following: MQCD results were based on studying 
the M5 brane obtained by lifting a brane
configuration with D4 branes between two
orthogonal NS5 branes. The gauge theory of the latter is $U(N)$ if we have
N D4 branes whereas the gauge theory on the M5 brane is $SU(N)$ and the 
 two coincide in the extreme IR because the
$U(1)$ is asymptotically free and decouples, but away from IR the field theory 
on the M5 brane is $SU(N)$. We can then ask
what happens to the $U(1)$ group $-$ do we 
lose all the information about it or can we recover the information 
in some limits? To answer that, we will inspire ourselves from the recent 
results about large N dualities of \cite{vafa}.

Recently, important results have been obtained concerning  field theories
on partially wrapped D-branes over non-trivial cycles of non-compact geometries.
In particular, based on the Chern-Simons/topological strings duality 
\cite{gova}, Vafa
suggested that D-branes wrapped over cycles have a dual description
where the D-branes have
disappeared and have been replaced by fluxes after transition in geometry 
\cite{vafa}.
This duality can be explained as a geometric flop in M theory on a $\G_2$
holonomy manifold \cite{amv,achar}. Other developments related to this
duality in 10 and 11 dimensions were discussed in
\cite{achar1,gomis,civ,en, Kachru,kaste,minakl,malnas,agavafa} 
and in the presence of orientifold planes
(besides the D-branes) in \cite{sv,eot}.
Vafa's duality considers strings and branes at conifold
singularities which have been studied
 extensively in the recent years, starting
with the work of Klebanov-Witten \cite{kw} and generalized to include
quotient conifold singularities \cite{mp,karch,Das}, nonconformal field theories
 \cite{kn,ot2} and theories with non-commutative moduli spaces \cite{ot3}.

We will be able to motivate, from the discussions related to the
 brane configurations for conifolds,
the important result that for N D5 branes on
$\P^1$ cycle of a resolved conifold, in the limit when 
the cycle is of very small size there is a transition from the
$U(N)$ theories on the branes to $U(1)$ theory. This is because 
there is a T-dual picture for the
resolved conifold which implies a configuration with N D4 branes on an interval 
between two orthogonal
NS5 branes which can be lifted to a {\it single}
M5 brane and is the starting point of the MQCD discussion.
What happens in M theory when the size of the $\P^1$ becomes small? The 
result, as we shall show, is that in this case the
curve M5 brane splits into a set of N ``plane'' M5 branes located at a  
specific point in the $x^{10}$ direction.
Also now on each planar M5 branes we have a
${\cal N} = 1, U(1)$ field theory. From where 
are the $U(1)$ groups coming from? They come exactly from the
$U(1)$ which was decoupled in 
the initial step of going from $U(N)$ to $SU(N)$. As explained in
\cite{WittenMQCD}, the $U(1)$ were 
decoupled because it had an infinite kinetic energy. What we show now
is that in the
limit of small size $\P^1$ cycle, the energy becomes finite so it makes sense to 
reconsider the $U(1)$ groups.

This also tells us that in the limit of small $\P^1$ cycle, one 
has to reconsider the issue of confinement and chiral
symmetry breaking. In \cite{WittenMQCD}, the issue was treated 
 for pure QCD and the result was that different
vacua were obtained for different curved M5 branes, the theory had
 a mass gap and the N vacua  rotated by
$Z_{2 N}$. In our case, we have condensation but
still have massless particle after condensation because we have the $U(1)$ 
groups on the plane M5 branes and the
group $Z_{2 N}$ rotates different $U(1)$ theories located on the different plane 
M5 branes.
We also obtain an interpretation of the domain walls and QCD strings in the
 small $\P^1$ limit.

We can now return and ask if our transition can be seen as an M theory lift of
the large N geometrical transition. This could be an alternative to the one 
involving $G_2$ manifolds \cite{amv,achar}.
We see that the plane M5 branes reduce to a type IIA configuration which is 
T-dual to the deformed conifold with
flux through the $\S^3$ cycle. The fluxes are due to bending of the 
 planar M5 branes along the $x^{10}$ direction implying a change in metric.
In 10 dimensions this reduces to a configuration of two intersecting 
NS5 branes with a two-form flux  whose T-dual is our required 
configuration.

\section{T-dual description of Vafa's ${\cal N}=1$ Duality}
In type IIA string theory, a four dimensional $\CN = 1$, $U(N)$ supersymmetric 
gauge theory is obtained by
wrapping $N$ D6 branes on the $\S^3$ in a complex deformed conifold (we
will simply call it a deformed conifold):
 \bea \label{rescon} xy-uv-\mu =0 \eea which is isomorphic to $T^*\S^3$ as a 
symplectic manifold after the rotation of symplectic
structure by the phase of $\mu$. In \cite{vafa}, Vafa has proposed a duality, in 
the large $N$ limit, between
this theory and type IIA superstrings without $D$--branes propagating on a 
K\"ahler deformed conifold which is a
small resolution of the conifold (we will call it a resolved conifold). Recall 
that the resolved
conifold is obtained by replacing the singular point of the conifold by a rigid 
$\P^1$ with normal bundle $\CO
(-1) + \CO(-1)$. By the definition of $\P^1$, each point $p$ of $\P^1$ 
represents a complex line $L_p$ and
$\CO(-1)$ (a.k.a. the tautological line bundle) over $\P^1$ is the complex line 
bundle on $\P^1$ with fiber $L_p$
over $p\in \P^1$, and $\CO(-1) + \CO(-1)$ is the direct sum of two copies
of
the $\CO(-1)$.  Moreover, it can be seen from the
toric description of the resolved conifold that the resolved conifold is the 
same as the total
space of the bundle $\CO(-1) + \CO(-1)$ over $\P^1$.

The large $N$ duality, which emerges from the embedding of the large $N$ 
Chern--Simons/~topological string duality of
Gopakumar and Vafa \cite{gova} in ordinary superstrings, states that  $\CN =1$ 
$U(N)$ theory on the deformed
conifold is dual to type IIA theory on the resolved conifold. In this duality, 
the branes disappear and are replaced
by $N$ units of RR flux through $\P^1$, and NS flux through the fibers of the 
normal bundle $\CO(-1)+\CO(-1)$. The complexified K\"ahler parameter t of $\P^1$ 
is related to the volume V of the $\S^3$ and
the string coupling constant by: \bea (e^t - 1)^N = \mbox{exp} (- V/g_s) \eea We 
note that when $\S^3$ has a
large volume V and when the 't Hooft coupling $N g_{YM}^2$ is small, the 
K\"ahler parameter $t \rightarrow 0$ and
the good description is the one with wrapped D-branes, whose worldvolume 
decouples from the bulk and we use the
open strings ending on the D-branes. When the K\"ahler parameter $t >> 0$, the 
blown-up description is good and
the wrapped D-brane picture is bad because then the volume of $\S^3$ should be 
negative. In this case we use the
close strings whose low-energy sector is given by supergravity. The $SU(N)$ 
gauge theory decouples from the bulk
when the size $t$ of the blown-up $\P^1$ is small and $t$ is identified with the 
glueball superfield $S =
-\frac{1}{32 \pi^2} TrW_\alpha W^\alpha$ of the $SU(N)$ theory, its expectation 
value corresponding to gaugino
condensation in the $SU(N)$ theory.

In the mirror description, the $U(N)$ theory is obtained from type IIB D5
wrapped on the rigid $\P^1$ in the resolved conifold and, in the
large N limit,
this is equivalent to type IIB on the deformed
conifold (\ref{rescon}) with RR flux on the $S^3$.
The deformation parameter $\mu$ will be identified with the $SU(N)$
glueball superfield $S$. Rather than the $N$
original D5 branes, there are now $N$ units of RR flux through $S^3$, and also 
some NS flux through the fiber of
the cotangent bundle $T^*\S^3$.
Moreover, by integrating the holomorphic
three form over the three cycle $\S^3$ and its dual cycle in $T^*\S^3$, one
obtains the superpotential 
\bea \We = S\log \left(\Lambda^{3N}/S^N \right)+ NS,
\eea and the $N$  vacua of $SU(N)$ $\CN =1$ supersymmetric Yang-Mills:
\bea
<S> = \exp (2 \pi i k /N) \Lambda^3.
\eea

Hence the large N duality arises from a
geometric transition from the resolved conifold to the deformed conifold.
We now take T dual of this geometric transition and later we will consider
the lifting to the M-theory.
We begin by  introducing  a circle action on the conifold and extend it to the 
resolved conifold and the deformed
conifold in a compatible manner.  Consider an action  $S_c$ on the conifold 
$xy-uv=0$: \bea \label{sc}
S_c:~~(e^{i\theta}, x) \to x ,~~ (e^{i\theta}, y) \to y~~(e^{i\theta}, u) \to 
e^{i\theta} u ,~~ (e^{i\theta}, v) \to
e^{-i\theta} v,~~ \eea The orbits of the action $S_c$ degenerates along the 
union of two intersecting complex
lines $y=u=v=0$ and $x=u=v=0$ on the conifold. Now, if we take a T-dual along
the direction of the orbits of the action, there will be NS branes along these
degeneracy loci as argued in \cite{bvs}. So we have two NS branes which are
spaced along $x$ (i.e. $y=u=v=0$) and $y$  directions (i.e. $x=u=v=0$)
together with non-compact direction along the Minkowski space
 which will be denoted by $NS_x$ and $NS_y$.  

 This action can be lifted to the resolved
conifold.
 To do that, we
consider two copies of $\C^3$ with coordinates $Z, X, Y$ (resp. $Z', X', Y'$) 
for the first (resp. second) $\C^3$.
Then $\CO(-1) + \CO (-1)$ over $\P^1$ is obtained by gluing two copies of $\C^3$ 
with the identification: \bea
\label{-1-1} Z' = \frac{1}{Z} ~, ~\quad X'= XZ ~, ~\quad Y' = YZ~. \eea The $Z$ 
(resp. $Z'$) is a coordinate of
$\P^1$ in the first (resp. second) $\C^3$ and others are the coordinates of
the fiber directions. The blown-down map from the resolved conifold $\C^3 \cup 
\C^3$ to the
conifold $\CC$ is given by \bea x=X=X'Z', ~~y=ZY=Y',~~u = ZX = X', ~~v= Y = 
Z'Y'. \eea From this map, one can see
that the following action $S_r$ on the resolved conifold is an extension of the 
action $S_c$ (\ref{sc}): \bea
S_r:~~(e^{i \theta }, Z) \to e^{i \theta} Z, ~~(e^{i\theta}, X) \to X,~~ 
(e^{i\theta},
Y) \to e^{-i\theta}Y\nonumber \\
 (e^{i \theta }, Z') \to e^{-i \theta} Z', ~~(e^{i\theta}, X') \to
e^{i\theta}X',~~ (e^{i\theta}, Y') \to Y' \label{sk} \eea 
The orbits degenerates along the union of two complex lines
$Z=Y=0$ in the first copy of $\C^3$ and $Z'=Y'=0$  in the second copy of $\C^3$.
Note that these two lines do not intersect and in fact they are separated by the 
size of
$\P^1$. 
Now we take T-dual along the orbits of $S_r$ of type IIB theory obtained
by wrapping $N$ D5 branes on the rigid $\P^1$. Again there will be two NS branes 
along the degeneracy loci of the action: one NS brane, denoted by
$NS_X$, spaced along $X$ direction (which is defined  by $Z=Y=0$ in the
first $\C^3$)
 and the other NS brane, denoted by $NS_{Y'}$  along
$Y'$ direction (which is defined by $Z'=X'=0$ in the second $\C^3$). Therefore 
the T-dual picture will be a brane configuration of D4 brane along the
interval with two NS branes in the `orthogonal' direction at the ends of
the the interval. Here the length of the interval is the same as the
size of the rigid $\P^1$.  As the rigid $\P^1$ shrinks to zero, the size of the 
interval goes to zero
and $NS_X$ (resp. $NS_{Y'}$) approaches to $NS_x$ (resp. $NS_y$) of the 
conifold.

Finally we will provide a circle action of the deformed conifold and a T dual 
picture under this action. Consider the
following circle action $S_d$ \bea \label{sd} (e^{i\theta}, x) \to x ,~~ 
(e^{i\theta}, y) \to y,~~ (e^{i\theta},
u) \to e^{i\theta} u ,~~ (e^{i\theta}, v) \to e^{-i\theta} v ,~~ \eea on the 
deformed conifold \bea xy -uv =\mu
\eea Then $S_d$ is clearly the extension of $S_k$ (\ref{sk}) and the orbits of 
the action degenerate along a
complex curve  $u=v=0$ on the deformed conifold. If we take a T-dual of the 
deformed conifold along the orbits of
$S_k$, we obtain a NS brane along the curve $u=v=0$ with non-compact direction 
in the Minkowski space which is
given by \bea \label{dns} xy = \mu \eea in the x-y plane.
Topologically, the curve (\ref{dns}) is $\R^1 \times \S^1$.
In the T-dual picture, the large N duality is achieved via  a transition from 
the brane
configuration of N coincident D4 branes between two `orthogonal' NS branes to
the brane configuration of a single NS brane wrapped on $\R^1 \times \S^1$
with gauge fields $A_{\mu}$ (which will be discussed in section 6).
\section{MQCD Transition} To investigate the large $N$ limit for a small $\P^1$, 
we appeal to Witten's MQCD
M5 brane of the brane configuration of the resolved conifold constructed in the 
previous section.

In MQCD \cite{WittenMQCD}, the classical type IIA brane configuration turns into 
a single fivebrane whose
world-volume is a product of the Minkowski space $\R^{1,3}$ and a complex curve 
in a flat Calabi-Yau manifold
\bea \label{M}M = \C^2 \times \C^*. \eea 
Recall that the T-dual brane configuration  of $N$ D5 branes wrapped on the 
rigid $\P^1$ in the resolved conifold
is  the N  D4 branes on the interval together with two NS branes $NS_X, NS_{Y'}$ 
at the ends. We denote the
coordinate of the interval by $x^7$ and the angular coordinate of the circle 
$\S^1$ in the 11-th dimension
 by $x^{10}$. After we combine them into a
complex coordinate \bea t = \exp ( -R^{-1}x^7 - i x^{10})\eea where $R$ is the 
radius of the circle $\S^1$ in the
11-th dimension, the world-volume of the M-theory fivebrane (a.k.a. M5 brane), 
corresponding to the brane
configuration of the resolved conifold, is given $R^{1,3} \times \Sigma$ where 
$\Sigma$ is a complex curve defined
by, up to an undetermined constant $\zeta$ \bea \label{m5res} y = \zeta x^{-1}, 
~~t = x^{N}\eea in $M$
 Here we are using $x, y$ instead of $X, Y'$
anticipating the identification after the transition and we will call $\Sigma$ a 
M5 curve. The equations
(\ref{m5res}) should be understood as an embedding of the punctured $x$ plane 
$\C^{*}$ into the Calabi-Yau space
$M$ by the map \bea \label{m5map} \C^{*} \longrightarrow \Sigma \subset M, 
~~~~~x \to (x, \zeta x^{-1}, x^{N})
\eea Hence $\Sigma$ is a rational curve and wraps around the punctured $t$ plane 
$\C^{*} \subset M$ $N$ times
which reflects the fact that there are $N$ coincident D4 branes along the $x^7$ 
direction in Type IIA.

Now if we consider the limit where the size of $\P^1$ goes to zero, then the 
$x^7$ direction in the M-theory will
be very small and negligible. The modulus of $t$ on $\Sigma$ will be fixed i.e. 
$\Sigma$ will be a curve in the
cylinder $\S^1 \times \C^2$ where $\S^1$ is the circle in the 11-th dimension 
and $\C^2$ are coordinatized by $x,
y$. In fact, the value of $t$ on $\Sigma$ must be constant because
 $\Sigma$ is holomorphic and there is no non-constant holomorphic map into 
$\S^1$. The holomorphicity is required
because of supersymmetry. Therefore the  M5 curve make a transition
 from a ``space'' curve into a ``plane'' curve. From (\ref{m5res}), we 
obtain two relation on $t$ and $t^{-1}$
 \bea \label{rel} t= x^N,~~ t^{-1} = \zeta^{-N} y^N.
 \eea
So there are $N$ possible plane curves which the M5 space curve $\Sigma$ can be 
reduced to: \bea \label{m5plane}
\Sigma_k:~~~t=t_0, ~~xy = \zeta \exp {2\pi i k/N}, \quad k =0, 1, \ldots , N-1. 
\eea Alternatively we may consider
these as $N$ possible relations between $x$ and $y$ on $\Sigma$ after 
eliminating $t$ because the dimension of
$t$ on $\Sigma$ is virtually zero and the information along $t$ is not reliable.

Since this is the limit where $g_sN$ is big, this degenerate M5 brane should not 
be considered as a M theory lift
of D branes. In this limit, the gravitationally deformed background without the 
D-branes is the right description
and this is a closed string geometric background. If one looks at the T dual 
picture of the deformed conifold
(\ref{dns}), this is exactly M theory lift of the NS brane of the deformed 
conifold! The size of $\S^3$ on the
deformed conifold depends on the expectation value for the gluino condensation 
and, for each value of the gluino
condensate we will have a different flux through the $\S^3$ cycle. We may 
intuitively consider the plane M5 as
one obtained from two intersecting M5 branes by smearing out the intersection 
point due to the flux from the
vanished D4 branes wrapped on  $\P^1$.

The  plane M5 branes (\ref{m5plane}) describe $N$ different $U(1)$ 
theories with  fluxes related to each
other by $\exp (2\pi i/N)$ after the $SU(N)$ group gets a mass gap. These $N$ 
different $U(1)$'s are just $N$
different vacua for the same theory, each obtained for a specific choice of 
$<S>$.
 The vacuum expectation values of the glueball superfield $S$ are then read
to be: \bea <S> = \exp (2\pi ik/N) \zeta \eea which agrees with the field theory 
result since $\zeta \sim N
\Lambda^3$ and also with \cite{hoo,strom,tv}. The $\CN =2$ vector multiplet 
consists of a
neutral $\CN =1$ chiral superfield and $\CN =1$ photon. The $\CN =1$ chiral 
superfield gets a mass due to the
presence of fluxes and is identified with $S$. So in MQCD, the large N duality 
occurs via a geometric transition
from the space M5 curve  to the plane M5 curve. The 11-dimensional supergravity 
solutions for this plane M5 curve
have been studied in ~\cite{bfhs}.

Before discussing the gauge theory on the plane M5 branes, let us
 consider the NS 
3-form flux which goes through the 
non-compact cycle dual to the  compact $\S^3$
cycle of the deformed conifold. After the T-duality, the $\S^3$ cycle 
translates into the waist of the NS brane $xy=\zeta$, considered as a
hyperboloid. The NS flux will be given by an integral of the holomorphic
1 form $dz/z$ over a noncompact 1-cycle dual to  the waist.
As in \cite{vafa}, the NS flux controls the size of the cycle that the 
D5 branes are wrapped on, in our case the cycle being $\P^1$. The
cycle $\P^1$ is a rigid one, this being the necesary condition to turn a NS
flux in the geometry. The size of the $\P^1$ cycle is related to the size of
the interval direction of the D4 branes which is related to the gauge coupling
constant of the $SU(N)$ theory. Therefore, the NS 
flux controls the magnitude of
the coupling constant of the  $SU(N)$ theory.

\section{Gauge theory on M5 brane} It has been shown that M5 brane (\ref{m5res}) 
gives rise to $SU(N)$ gauge
theory and $U(1)$ gauge fields are 
decoupled from the theory \cite{WittenMQCD}. 
It is puzzling to see the massless
$U(1)$ photons after the transition.
 To understand this phenomenon better, we recall the low
energy effective four-dimensional theory from M5 brane derived in \cite{wit}. 
Consider in general a fivebrane
whose world-volume is $\R^{1,3}\times \overline{\Sigma}$ where 
$\overline{\Sigma}$ is a compact Riemann surface of
genus $g$. According to \cite{klmvw, verlinde}, in the effective 
four-dimensional description, the zero modes of the
antisymmetric tensor give $g$ Abelian gauge fields on $\R^4$. The coupling 
constants and theta parameters of the
$g$ Abelian gauge fields are described by a rank $g$ Abelian variety which is 
the Jacobian
$J(\overline{\Sigma})$. Let \bea T = F\wedge \Lambda + \ast F \wedge \ast 
\Lambda, \eea where $F$ is a two-form
on $\R^4$, $\Lambda$ is a one-form on $\overline{\Sigma}$, and $\ast$ is the 
Hodge star. This $T$ is self dual,
and the equation of motion $dT = 0$ gives Maxwell's equations $dF = d \ast F =0$ 
along with the equations
$d\Lambda = d\ast \Lambda = 0$ for $\Lambda$. So $\Lambda$ is harmonic 
one-form and every choice of a harmonic
one-form $\Lambda$ gives a way of embedding solutions of Maxwell's equations on 
$\R^4$ as a solution of the
equations for the self-dual three-form $T$.

We now consider $n+1$ parallel Type IIA NS branes joined by D4 branes. The M5 
brane world-volume will be of the
form $\R^{1,3} \times \Sigma$. The $\Sigma$ is obtained by joining $n+1$ copies 
of $\C$ by $N_k$ tubes $\C^{*}$
between $k$-th and $(k+1)$-th $\C$. By adding $(n+1)$ points, the $\Sigma$ can 
be compactified into a compact
Riemann surface $\overline{\Sigma}$ of genus $g =\sum_{k=1}^n (N_k -1)$ as was 
shown in \cite{wit}. A Jacobian
(or quasi-Albanese)  of a non-compact Riemann surface can be defined in terms of 
its mixed Hodge structure
\cite{deligne}. For the Riemann surface $\Sigma$, the Jacobian $J(\Sigma)$ fits 
into an exact sequence of
algebraic groups: \bea \label{jac}  1\longrightarrow (\C^*)^n \longrightarrow 
J(\Sigma ) \longrightarrow
J(\overline{\Sigma}) \longrightarrow 0 \eea where $J(\overline{\Sigma})$ is the 
usual Jacobian of the
compactification $\overline{\Sigma}$ of $\Sigma$. This exact sequence does not 
split in general as the Jacobian
$J(\Sigma)$ represents a non-trivial elements of 
$\mbox{Ext}(J(\overline{\Sigma}), (\C^*)^n ).$ In the effective
action for the four dimensional gauge theory, the harmonic forms from the 
non-compact part $(\C^*)^n$ decouples
from the theory because the corresponding M5 brane kinetic energy $\int_{R^{1,3} 
\times \Sigma} |T|^2$ becomes
infinity as the non-trivial harmonic one form $d\log \,z$ on $\C^*$ is not 
square-integrable~\cite{wit}. Thus the
low energy effective action is determined by $J(\overline{\Sigma})$ and the 
gauge group is
$\prod_{k=1}^{n}SU(N_k)$. On the other hand, according to the large N duality 
proposal of \cite{civ}, the gauge
theory becomes $U(1)^n$ after the transition and it agrees with $U(1)^n$ theory  
from $U(N)$ theory after the
$SU(N_i)$ gets a mass gap and confine in the breaking \bea U(N) \to 
\prod_{k=1}^n U(N_k).\eea Hence after the
transition, the coupling from the non-compact part $(\C^*)^n$ of the Jacobian 
$J(\Sigma)$ has been restored. From
the M-theory point of view, the large N duality is a transition from the compact 
part $J(\overline{\Sigma})$ to
the non-compact part $(C^*)^n$ of the Jacobian $J(\Sigma)$. So there will be a 
duality between $\prod_{k=1}^n
SU(N_k)$ and $U(1)^n$ and  the former will be described by the space M5 curve 
and the latter will be described by
the plane M5 curve. In this paper we will deal with the simplest
case of $n=1$ and more complicated case ($n>1$) will be dealt in 
\cite{toappear}.
As the $\P^1$ gets smaller, there will be gaugino condensation in 
$SU(N_k)$ theory, the decoupled $U(1)$ will be restored back into the picture.
 
To see this more precisely, let us consider
the case of  N D5 brane wrapping $\P^1$ in the resolved conifold for simplicity 
(the arguments for the general
case is similar).  After the transition, the plane M5 curve 
will be given by $\Sigma:  ~xy = \zeta$ in the x-y plane which
is $\C^*$. But it is wrong to use the flat metric for $\C^*$ here. This M5 was 
obtained from two NS, say $NS_x,
NS_y$, branes by collapsing the $\P^1$ cycle. 
Hence for large values of $x$, the metric on M5 curve will be like
that of  $NS_x$ and for small values of $x$, the metric on M5 curve will be 
like that of $NS_y$ at large value
of $y$ since two NS branes are 
smeared out at the intersection point. The metric on 
M5 is nothing but the induced
metric from the flat metric on the x-y plane which is the 11 dimensional
metric in our M theory. 

Even though for  such a
metric, the integral $  \int_{\Sigma} d\log \,x \wedge \ast d\log \,x $ is
still divergent, now we can regulate the infinity to obtain finite period 
by putting a cutoff, thus using the same argument for cutoff as in \cite{civ}.
Because $|\zeta|$ has dimension 3, $|x|$ and $|y|$  have dimensions $3/2$. 
By putting
the cutoff at $|x| = \Lambda_0^{3/2}$, the integral 
\bea 
\label{cutoff}
\int_{\Sigma} d\log \,x \wedge \ast d\log \,x& \sim &
2\int_{\Lambda_0^{3/2}>
|x|
>|\zeta|^{1/2}} dx\wedge d\bar{x}/ |x|^2   \\ \nonumber
&=& 3 \log \Lambda_0 - \log |\zeta|
\eea 
becomes finite and these harmonic one forms are exactly captured by the 
non-compact
part $(\C^*)^n$ of the Jacobian $J(\Sigma)$. We observe that the IR divergency 
of the integral is removed because of $|\zeta|$ so the only necessary 
cutoff is
in the UV. 

Hence the decoupled $U(1)$ gauge fields will be restored after the
transition. In view of this, it seems to be reasonable to assume that there is a 
MQCD theory associate to the
Jacobian $J(\Sigma)$ which is a $\prod_{k=1}^nU(N_k)$ gauge theory and is not 
broken to $U(1)^n \times
\prod_{k=1}^n SU(N_k)$ since the exact sequence (\ref{jac}) does not split in 
general.

{}From (\ref{cutoff}) we can extract the coupling constant for the theory as:
\bea
\frac{1}{g^2} \sim 3 \log \Lambda_0 - \log |\zeta|
\eea
which is the expected running of the coupling constant for 
$D = 4, {\cal N} = 1$ Yang-Mills theory if we replace 
the cutoff $\Lambda_0$ by
the scale of the gauge theory, denoted by $\Lambda$. The coupling constant of 
$U(1)$ theory is $1/N$ of the coupling constant for $U(N)$ theory.

There is also a discrete $\Z_2$ symmetry of exchanging $x$ and $y$ \bea x \to y, 
~~y\to x~~, t \to \zeta^N t^{-1}
\eea reverses the orientation of strings stretched between different D4 branes. 
This $\Z_2$ symmetry corresponds
to the exchange of two factors of $\CO(-1)$ in the $\CO(-1) +\CO(-1)$ over 
$\P^1$ combined with the involution $w
\to w'$ on $\P^1$ which is an exact symmetry of the vacuum.

\section{Domain Walls and QCD Strings}
Confinement is one of the most mysterious and puzzling aspects of QCD.  Assuming 
that $\CN =1$ MQCD is in the
same universality class as QCD \cite{WittenMQCD}, we can
study some aspects of QCD strings.  As a consequence of having $N$ different 
vacua
after the transition, there can be domain walls separating different vacua. 
In the
large $N$ limit, they behave as M5 branes where the QCD strings can 
end~\cite{WittenMQCD}.

Recall that the vacua are described by $N$ plane M5 curves: \bea 
\Sigma_k:~~~t=t_0, ~~xy = \zeta \exp {2\pi i
k/N}, \quad k =0, 1, \ldots , N-1. \eea  A domain wall is a physical situation 
that for one region of $x^3$ looks
like one vacuum of the theory and for another region of $x^3$ looks like another 
vacuum. Here $x^3$ is one of the
three spatial coordinates in $\R^{1,3}$ and the physics should be independent of 
the time $x^0$ and the other two
spatial coordinates $x^1, x^2$. For 
 convenience, we compactify $x^3$ direction 
into a circle $x^3$ from now one
and use the angular coordinate $\theta$ for the phase of $\S^1$. So we are 
assuming that we have the same physics at
$x^3 \to \infty$ and $x^3 \to -\infty$. In M theory, such a 
physical situation 
can be realized as a fivebrane that
interpolates between the N plane M5 curves $\Sigma_k$ describing different 
vacua. Therefore we consider a fivebrane of
the form $\R^3 \times \CD$ where $\R^3$ is parameterized by $x^0, x^1, x^2, $ 
and $\CD$ is a three-surface in the
seven manifold $\tilde{M} = \S^1 \times M$ where $\S^1$ is parameterized by
$\theta, ~~0 \leq \theta < 2\pi$. We define $\CD$ in
$\tilde{M}$ with fixed $t$ as
\bea t = t_0, ~~xy = \zeta \exp (i \theta ). \eea
We have a smooth family of holomorphic curves
parameterized by $\theta$ in $\tilde{M}$.  Since it is locally diffeomorphic
to a product of $\R$ and a Riemann surface, it is supersymmetric. For $\theta = 
2 \pi k/N, k = 0, \ldots , {N-1}$, the three-surface $\CD$
will be a holomorphic curve $\Sigma_k$ which describes the $\CN =1$ $U(1)$ 
supersymmetric Yang Mills with vacua
of the glueball superfield $S$ of the $SU(N)$ theory. Hence this is the domain 
wall connecting $\CN =1$ $U(1)$
theory with all possible different vacua.

Recall that the vacua are described by $N$ plane M5 curves $\Sigma_k$:
 \bea t =t_0,~~ xy = \zeta \exp (2\pi i k/N), ~~ k=0, \ldots , N-1\eea in $M$.
We now define QCD strings as open oriented onebrane $C_k$ in $M$ with fixed $t$: 
\bea t&=&t_0,\\
\nonumber
 x&= &t_0^{1/N} \exp (2\pi i (k+\sigma )/N),\\
\nonumber y&=&\zeta t_0^{-1/N}. \eea The curve  $C_k$  connects two vacua 
$\Sigma_k$ and $\Sigma_{k+1}$.
 As a topological object,
they are classified as the element of a relative homology $H_1(M, \cup\Sigma_k, 
\Z)$. The relative homology fits
into a long exact sequence: \bea \ldots \to  H_1 (\cup\Sigma_k,\Z) \to H_1(M,\Z) 
\to H_1(M, \cup\Sigma_k , \Z) {\to}
H_0(\cup\Sigma_k, \Z)\to \ldots \eea  Using this sequence and topological 
properties of $\Sigma_k$ and $Y$, it can be
shown that \bea H_1(M, \cup\Sigma_k , \Z) \cong \Z^n. \eea The $N$ of $C_k$ will 
annihilate as an element of $H_1(M,
\Z)$ i.e. they can be detached from $\cup \Sigma_k$ as the boundaries add up to 
zero and will form a long closed string
in $M$ and can be deformed into a point. By construction, all of the QCD strings 
lie entirely in the domain wall
$\CD$ and hence stable. 

Recently, the BPS domain wall of supersymmetric Yang-Mills theory for
arbitrary gauge theory has been
studied~\cite{archavafa}. For $SU(N)$ group, the M-theory formulation on $\G_2$
holonomy  geometry has been used. It would be desirable to see the results
of
\cite{archavafa} from the MQCD set-up as above.
\section{Brane Construction from Fluxes}

So far we've studied the large N transition purely from M-theory point of 
view. What happens in string theory?
In this section we shall provide an intuitive understanding of the 
large $N$ duality using type IIA brane constructions. But before 
we dwell into that we need some more 
details on the conifold.
 As was shown in the previous sections, the T dual of  type IIB theory on a
 conifold  can be regarded as
two NS5 branes at a point $x^6= x^7= 0$. The 
distance between the two NS5 branes along $x^7$ is given by 
the resolution of a
conifold and the distance along
$x^6$ is
given by the value of $B_{NSNS}$ field 
on the vanishing two cycles of the base 
sphere in the T-dual model. The latter can be shown easily if we have some D4
branes along the compact direction $x^6$ and also stretched along 
$x^{0,1,2,3}$. The distance between the NS5 branes determines the $3+1 ~d$ 
gauge coupling which in turn is determined by the $\int B_{NSNS}$ over the
two cycle $S^2$ of the base of the conifold. If we have both the situation,
namely, a resolved conifold with some $B_{NSNS}$ flux through the two 
cycles then the T-dual picture is given by two NS5 branes separated along
a distance $z_0$ in a complex $z=x^6+ix^7$ plane.

The $S^3$ of the base is a Hopf fibration of a circle on a sphere $S^2$. The 
circle, whose coordinates we specify
as $\psi$ is non trivially twisted over the base $S^2$. This twist can be seen 
to come from the NS5 brane charges
in the T-dual picture. To make this connection precise we first make an 
identification of the brane coordinates
$x^n$ with the conifold coordinates $\theta_i,\phi_j,\psi$. For simplicity we
can have the following
identifications: \bea \label{identif} dx^{4,8}\to sin~\theta_{1,2}~d\phi_{1,2}, 
~~~~~ dx^{5,9}\to
d\theta_{1,2}\eea

In these units the magnitude of the two $B_{NSNS}$ fields $B_i$ 
satisfy\footnote{There are various $B$-fields in the model. The $B_i$'s 
discussed
here are the sources of the two NS5 branes. We take $B_1\equiv B_{64}$ and 
$B_2 \equiv B_{68}$ only as it is easy to show that the other components
can be gauged away. The $\int B_{NSNS}$ field discussed above is a flux 
through the two cycle.}
 the 
equation \bea
\label{Bfieldeq}{\frac{1}{sin\theta_i}}~ {\frac{\del}{\del\theta_i}}~ 
(sin\theta_i~B_i) = constant \eea whose
solutions are given by $cot~\theta_1$ and $cot~\theta_2$. Thus the non trivial 
twist of the cycle $\psi$ on the
base is governed by the metric \bea \label{psimetric} ds^2 = 
(d\psi+cos~\theta_1~d\phi_1+
cos~\theta_2~d\phi_2)^2  \eea Here $\psi = x^6$. There are also two $S^2_i, 
~i=1,2$ whose metric is given by the
usual polar coordinates \bea \label{stwo} 
\sum_{i=1}^2~~(d\theta_i^2+sin^2\theta_i~d\phi_i^2) \eea However it
turns out that in homology, the $S^2$ factor in $S^2\times S^3$ is the 
difference of the two $S^2$'s parameterized
by $\theta_1,\phi_1$ and
 $\theta_2,\phi_2$ respectively. To see this observe that we can write two
2-forms \bea \label{twoforms} sin~\theta_1~ d\theta_1~ d\phi_1 \pm sin~\theta_2 
~d\theta_2~ d\phi_2 \eea both of
which live on the two $S^2$ factors and are independent of the $U(1)$ fiber.

Both of these 2-forms can be written formally as exact forms, namely the above 
expressions are equal to \bea
\label{formsareeq} d(cos~\theta_1~d\phi_1 \pm cos~\theta_2~d\phi_2) \eea 
However, the expressions in brackets
above are ill-defined when any of the $\theta_i$ is equal to $0$ or $\pi$, since 
in that case we are at the north
or south pole of one of the 2-spheres and the coordinate $\phi_i$ is undefined 
there. However, the term in the
$+$ sign can be modified to: \bea \label{formmod} d(d\psi + cos~\theta_1~d\phi_1 
+
 cos~\theta_2~d\phi_2) \sim de^{\psi} \eea
where $e^{\psi}$ is one of the five vielbeins given in appendix (A) of Ref. 
\cite{candelas}, and is globally
defined because $\psi$ is allowed to have gauge transformations. It follows that 
the term with the $+$ sign in
eq. (\ref{twoforms}) is genuinely exact, leaving the one with the minus sign as 
the representative of the second
cohomology. From this the above claim follows.

Now, with the above back-ground,
let us try to understand this duality from our brane construction. However 
there is a small subtlety here. The brane construction that we use here for
deformed conifold is an approximation because the derivation relies on the
fact that some of the directions are actually de-localized. Therefore the 
brane construction for deformed conifold, in this section, should be regarded
as for de-localized branes. The exact brane construction is been derived 
in sec. 2 which takes only localized branes. But since for completely
localized branes we do not know how to calculate the exact supergravity 
metric we, for the purpose of this section only, take the approximate 
framework. Also, as it will be clear soon, we need only the brane 
construction for the conifold with $S^3 \to 0$.
   
We start with the deformed conifold with RR
flux on the $S^3$. For simplicity we shall assume that the total flux is 
constant. Now let us shrink the $S^3$
cycle to zero size. In the limit when the cycle has shrunk to zero size then the 
resulting manifold is a singular
variety and is our old friend conifold.

The T-dual of this process is known. As we discussed in sec. 2, the T-dual 
of a deformed conifold is a set of two intersecting NS5 branes on a 
curve $xy = \mu$ and in the limit $\mu \to 0$ 
it is nothing but two NS5 branes intersecting at a point (which is, of
course, a $3+1$ dimensional plane). 
But now this is not enough. Recall that in the IIB picture we have a large
$H_{RR}$ flux at the conifold point. The $H$ field is on $S^3$ and has 
components $H_{\psi \mu \nu}$ where
$\mu,\nu$ are coordinates on the base sphere. The T-dual of $H_{\psi \mu \nu}$ 
gives us $F_{\mu \nu}$ in type IIA
theory where $F = dA$, $A$ being the RR gauge field. Therefore we have  $N$ 
units of flux at the point where the
two NS5 branes meet. It is important to ask what directions do the $\mu,\nu$ 
coordinates span. The coordinates
$\mu,\nu$ are on the base of $S^3$. The cohomology of the $S^3$ is generated by 
the third cohomology
$H^3(T^{11})$ as: \bea \label{thirdcoho} e^{\psi}\wedge e^{\theta_1}\wedge 
e^{\phi_1} - e^{\psi}\wedge
e^{\theta_2}\wedge e^{\phi_2} \eea where we have already defined $e^{\psi}$ and 
$e^{\theta_i},e^{\phi_i}$ are
defined in terms of the conifold variables as: \bea \label{ethetai} 
e^{\theta_i}= d\theta_i, ~~~~~~~~e^{\phi_i}=
sin~\theta_i~d\phi_i, ~~i=1,2 \eea Therefore $\mu, \nu$ span all the four 
directions $x^{4,5,8,9}$ and hence $F$
will have components along both the spheres $x^{4,5}$ and $x^{8,9}$.

When the $S^3$ is shrunk to zero, we have, in the T-dual model, the two NS5 
branes at the point $x^6 = x^7 = 0$
and the directions $x^{4,5}$ and $x^{8,9}$ of the NS5 branes wrapped on 
vanishing spheres. This is a singular
configuration (with flux per unit area approaching infinity) and because of this 
a {\it four-brane} is created by
a mechanism similar to the  Hanany $-$ Witten effect\cite{hw}. To understand 
this let us recall some basic facts about the
NS5 brane.

On the world volume of a NS5 brane there propagates a chiral $(2,0)$ tensor 
multiplet whose fields are
$(B^+_{ij}, 5\phi)$. $B^+$ is an anti-self dual two forms on the NS5 brane whose 
coordinates are $\sigma^{i,j}$
and $\phi_i, i=1,..,5$ are the five scalars. Two out of these five scalars are 
periodic. A simple way to see this
would be as follows \cite{senkk}:

If we T-dualize orthogonal to a de-localized
 NS5 brane of type IIA we shall get a configuration of Kaluza-Klein (KK)
monopole in type IIB theory. As have been discussed in many previous works, the 
KK monopole supports a
normalizable harmonic form which is anti self-dual. The four form of type IIB 
when reduced over this two-form
gives rise to the two form $B^+_{\mu\nu}$ of the NS5 brane. The two $B$ fields 
of type IIB $B_{RR}$ and
$B_{NSNS}$ give the two periodic scalars on the NS5 brane. The other three 
scalars come from the translational
zero modes of the KK monopole.

The background gauge field couples to the world volume of the NS5 brane through 
the following coupling: \bea
\label{couponns} \int ~ A \wedge *d\phi = \int ~ F \wedge C_4 \eea where $dC_4 = 
*d\phi$ is the six dimensional
dual of a world volume periodic scalar. From type IIB point of view this 
coupling can be motivated from the
kinetic term of the $H$ fields.

Because there is a background expectation value of the gauge field, this will 
induce a source $\int C_4$ on the
world-volume of the NS5 brane. A generic  source will break all supersymmetry on 
the world volume. The susy
preserving source will tell us that we have a {\it four-brane} stretched between 
the two NS5 branes.

Along what direction should this D4 stretch? It can be argued that the four- 
brane should be along $x^{0,1,2,3}$
and $x^7$. Observe that in our model there are two orthogonal directions $x^6$ 
and $x^7$. The limit where $S^3$
is of zero size is, as we discussed above, when the $x^{4,5}$ and $x^{8,9}$ 
spheres are of zero size and the
branes are on top of each other. In the type IIB T-dual model the conifold 
transition will tell us that we go to
a non-singular configuration by growing a {\it two-sphere}. The branes should 
therefore be now separated along
$x^7$ because this process is the T-dual of resolution in the conifold. Thus the 
final configuration will be a D4
stretched between two NS5 branes separated along $x^7$, which is precisely the 
configuration that we discussed in
detail in the previous sections. But this is nothing but the T-dual of a 
resolved conifold with a D5 wrapping the
$S^2$! Observe that we have arrived at this duality by a transition via a 
singular configuration $-$ a conifold
transition.

In the discussion above we gave a physical picture to motivate the 
large N duality of Vafa. One last thing is to see how from M-theory this
transition could be understood. Recall that in M-theory 
we start in ``reverse'' 
order, i.e we start with a configuration of a {\it single} M5 brane 
coming from a N D4 branes between two orthogonal NS5 branes in type IIA
 theory. After the transition the space M5 branes become N ``planar''
M5 branes which are located at a particular position in
$x^{10}$. What does an 
observer in $d=10$ sees? The two M5 which intersect at a point in 
$x^{10}$ will appear as two intersecting NS5 branes satisfying the 
equation $xy = m\zeta$ where $m$ is a constant factor. The bending along the 
$x^{10}$ direction, which gives non trivial components of the metric, will
appear as gauge fields $A_{\mu}$ in 10 dimensions. Again 
this is precisely the
brane configuration we started with in the beginning of this section. 

Finally, what about the $U(1)$ fields?
In the above discussions we have motivated the $U(1)$ fields as the one
coming from the $(2,0)$ fields of M5 branes. Also we know from the large
N duality that the $U(1)$ gauge fields have their
 origin from type IIB four form 
reduced over holomorphic three form of the Calabi-Yau. How are they related?
As we discussed in some
details the two form of M5 brane has its origin from the four form of type
IIB. Using various dualities of NS5 brane to a Taub-NUT space and 
two intersecting NS5 branes to a conifold one can easily show that the two
$U(1)$ are the same. This is how the transition from $U(N)$ theory on the
brane to $U(1)$ theory on the bulk can be seen.  

Before we end,
one very interesting direction is to study the interpolation between the 
supergravity solution corresponding to $D5$ branes wrapped on the $\P^1$ cycle 
of
the resolved conifold and the one corresponding to fluxes on the $\S^3$ cycle of 
the deformed conifold \footnote{We would like to thank 
Igor Klebanov for discussions 
concerning this supergravity interpolation.}. On one side 
we have the results of
\cite{ks,mn} for $D5$ branes wrapped on $\P^1$ and on the other side we have 
results 
concerning the deformed conifold without fluxes \cite{ohta,pando}. The work 
towards
an interpolating solution is in progress. 

\section{Acknowledgment}

We thank Rajesh Gopakumar, David Kutasov, Juan Maldacena, Carlos Nunez and 
Angel Uranga for helpful discussions.
We especially thank Igor Klebanov and Cumrun Vafa for comments on the
 manuscript.
The research of KD is supported by Department of Energy grant no. 
DE-FG02-90ER40542. The research of 
KO is supported by NSF grant PHY 9970664. The research of RT is supported by
DFG.

\end{document}